\documentclass{elsart}
\usepackage[dvips]{graphicx}
\begin{document}
\begin{frontmatter}
\title{Ensemble Inequivalence: A Formal Approach}

\author{F. Leyvraz\thanksref{francois}}
and
\author{S. Ruffo\thanksref{stefano}}
\address{Dipartimento di Energetica ``S. Stecco", 
Universit\'a di Firenze, Via S. Marta, 
3 I-50139, Firenze, Italy, INFM and INFN, Firenze}

\thanks[francois]{E-mail: leyvraz@ce.fis.unam.mx.\\ Permanent address: Centro de
Ciencias F\'\i sicas, Av. Universidad s/n, 62251, Cuernavaca, Morelos, Mexico}
\thanks[stefano]{E-mail:ruffo@avanzi.de.unifi.it}

\begin{abstract}
Ensemble inequivalence has been observed in several systems. 
In particular it has been recently shown 
that negative specific heat can arise in the microcanonical 
ensemble in the thermodynamic limit for systems with long-range interactions.
We display a connection between such behaviour and a 
mean-field like structure of the partition function. Since
short-range models cannot display this kind of behaviour,
this strongly suggests that such systems are necessarily non-mean
field in the sense indicated here. We further show that a broad
class of systems with non-integrable interactions are 
indeed of mean-field type in the sense specified, so that
they are expected to display ensemble inequivalence as
well as the peculiar behaviour described above in
the microcanonical ensemble. 
\end{abstract}
\end{frontmatter}
\vspace{-1.25truecm}

\newcommand\pb{\overline p}
\newcommand\qb{\overline q}
\newcommand\mass{{\Psi}}
\newcommand\taubar{\overline\tau}

\section{Introduction}
Particle or spin systems for which the pairwise interaction
potential decays at large distances with a power smaller than
space dimension are called {\it long-range} or {\it non-integrable}.
It has been suggested~\cite{Thirring,Gross} that at first order phase
transitions such systems should display ensemble inequivalence
also in the thermodynamic limit. A few examples where this is
explicitly shown have been published~\cite{Thirring,Antoni,Barre,Cohen}.
The specific heat, which is always positive in the canonical
ensemble, may become negative in the microcanonical, even if the
interaction is not singular at short distances like for 
gravity~\cite{Padmanabhan}. Moreover, temperature jumps at critical
energies may be present.
In this paper we present a formal approach within which all these
features can be put into context. The approach is based on the assumption,
which we justify afterwards, that a sort of Landau free energy can be
always introduced for long-range systems.
\section{Mean-field and Ensemble Inequivalence}
In this section, we show the main result of this paper. We say that
a system is of {\it mean-field type\/} if it satisfies the following
condition:
\begin{equation}
Z_c(\beta)=\int_{-\infty}^\infty \exp\left[-N\mass(\beta,m)\right]dm,
\label{eq:mean-field}
\end{equation} 
where $Z_c(\beta)$ is the canonical partition function and $\mass(\beta,m)$
is an analytic function of $\beta$ and $m$. Note that we require this
for finite values of $N$, but that we additionally require
the function $\mass(\beta,m)$ to be independent of $N$. 
Infinite range models always have this 
property, as follows from their solution.

From this we can derive that ensemble inequivalence 
will in general occur. Indeed, from (\ref{eq:mean-field}) one obtains
the following expression for the microcanonical partition function:
\begin{equation}
Z_m(\epsilon)=\int_{-\infty}^\infty dm\int_{-i\infty}^{i\infty}
\frac{d\lambda}{2\pi i}\exp\left[
N(\lambda\epsilon-\mass(\lambda,m)
\right].
\label{eq:double-integral}
\end{equation} 
One can now perform a saddle point integral to estimate the value
of the $\lambda$ integral for large values of $N$. We argue that the
dominant saddle point must lie on the real axis: If it were otherwise, 
we would have two
complex conjugate dominant saddle points, and hence an oscillatory
behaviour of the partition function, contradicting positivity
of $Z_c(\beta)$. We may therefore limit 
ourselves to the consideration of real saddle points. Since the 
integration path is perpendicular to the real axis, the real minima
of the argument of the exponential will correspond to allowed saddles, 
since these will correspond to local maxima when traversed along 
the imaginary axis. One is thus led to the following expression
for the microcanonical entropy  per particle:
\begin{equation}
S_m(\epsilon)=\max_m\min_\lambda[\lambda\epsilon-\mass(\lambda,m))].
\label{eq:entropy-micro}
\end{equation} 
This result can now be compared
with the standard result for the canonical ensemble:
\begin{equation}
S_c(\epsilon)=\min_\lambda\max_m[\lambda\epsilon-\mass(\lambda,m))].
\label{eq:entropy-canonical}
\end{equation} 
From these formulas, a few consequences are immediate:
\begin{enumerate}

\item The two forms of the entropy need not be equal. Indeed, if 
we could prove that $\mass(\lambda,m)$ has a unique extremum under
certain conditions, we could argue for equality. However, we 
know that this is not generally the case: Whenever a phase transition 
occurs, the function $\mass(\lambda,m)$ may have multiple
extrema at least as a function of $m$, thus precluding any simple
statements about the identity of the two entropies.

\item
Let us consider the behaviour of the function 
$\mass$ at its equilibrium points in the two ensembles. For the 
canonical problem we find the usual stability conditions
for the thermodynamic potential $\mass$. 
However, if one goes through the same computation for the
microcanonical ensemble, one finds
that concavity in $\lambda$ and indefiniteness of the Hessian
are sufficient for stability.
Note in particular how this allows the second derivative
of $\mass$ with respect to $m$ to take either sign. 

\item
The microcanonical entropy 
is a maximum over concave functions, which need not
be concave. Specifically, an explicit evaluation
of the second derivative of $S_m(\epsilon)$ yields
\begin{equation}
\frac{d^2S_m}{d\epsilon^2}=\frac{\partial^2\mass}{\partial\lambda^2}
-\frac{
(\partial^2\mass/\partial m\partial\lambda)^2
}{
\partial^2\mass/\partial m^2
}.
\label{eq:entropy-2-derivative}
\end{equation} 
Again, since $\partial^2\mass/\partial m^2$ can take both signs,
one sees that the specific heat can do so as well. 
In fact, one finds that 
the sign of $d^2S_m/d\epsilon^2$ is the opposite of that
of $\partial^2\mass/\partial m^2$. The specific heat
is thus negative exactly when the value of $m$ corresponding
to microcanonical equilibrium is unstable from the point
of view of the canonical ensemble. 

In the canonical case, it can also be shown using standard 
properties of the Legendre transform that the canonical entropy
(\ref{eq:entropy-canonical}) cannot have a discontinuous 
derivative. Again, it can be seen both from a general point of 
view and via explicit examples, that this reasoning does
not hold for the microcanonical entropy (\ref{eq:entropy-micro}).
This means that temperature can be a discontinuous function of
energy.

\item
Finally, we note that, since the equivalence between ensembles 
is rigorously proved for interactions which are of sufficiently
short-range, our arguments strongly suggest that for 
these systems such functions as $\mass$
do not have the required analyticity properties. The argument is not fully 
rigorous, but it is certainly very suggestive.
\end{enumerate}

\section{Models with Long-range Interactions}
In this Section, we show that a large class of models having non-integrable
interactions, actually satisfy our criterion (\ref{eq:mean-field}).
For definiteness' sake we restrict
ourselves to spin models on a lattice. 
Consider the following Hamiltonian
\begin{equation}
H\left[s(\vec i)\right]=L^{-(d-\alpha)}\sum_{\vec k, \vec l}
\frac{s(\vec k)s(\vec l)}{|\vec k-\vec l|^\alpha}+
\sum_{\vec k}V\left[s(\vec k)\right].
\label{eq:hamiltonian}
\end{equation} 
Here the $s$ are spins which run over a discrete set $S$, the indices $\vec k$
run over a $d$-dimensional lattice and $\alpha$ is an exponent between zero
and $d$. The normalization of the interaction by $L^{-(d-\alpha)}$
guarantees that the Hamiltonian is in fact {\it extensive}.

To express the partition function in the form (\ref{eq:mean-field}), 
we introduce coarse-grained variables $m(\vec x)$ for the magnetization. 
Using the long-range nature of the interaction, one
obtains an expression for the 
partition function in terms of a functional integral over all
$m(\vec x)$. This integral is found to be 
dominated by a single saddle-point, which corresponds to 
a function $m(\vec x)$ equal to a constant.  
One can therefore replace the functional integration
by an ordinary integration  over
this constant value and one has cast the partition
function in the form (\ref{eq:mean-field}). For details
about this derivation see Ref.~\cite{Leyvraz}.

\begin{ack}
We thank E.G.D. Cohen, J. Barr\'e, T. Dauxois, M.C. Firpo, D.H.E. Gross,
A. Torcini and E. Votyakov for stimulating discussions.
This work is supported in part by the MURST-COFIN00 project {\it Chaos
and localization in classical and quantum mechanics}
as well as by CONACyT project 32173-E and DGAPA IN112200.
\end{ack}

\end{document}